\documentclass[12pt]{article}
\usepackage{epsfig}

\newcommand{\uu}{\mbox{\boldmath $u$}}

\newcommand{\rr}{\mbox{\boldmath $r$}}

\newcommand{\q}{\mbox{\boldmath $q$}}
\newcommand{\PP}{\mbox{\boldmath $P$}}

\newcommand{\p}{\mbox{\boldmath $p$}}

\newcommand{\kk}{\mbox{\boldmath $k$}}

\newcommand{\htts}{\mbox{\boldmath$\hat{t}\kern1pt$}}

\begin{document}

\begin{center}
ANALYTICAL CALCULATION OF THE NEUTRONS SPECTRUM\\
FOR DIRECT MEASUREMENT OF N-N SCATTERING\\AT PULSED
REACTOR YAGUAR\\

V.~K.~Ignatovich\footnote{
e-mail ignatovi@nf.jinr.ru}\\

FLNP JINR Dubna 141980 RF
\end{center}

\begin{abstract}
Analytical calculation of a single neutron detector counts per
YAGUAR reactor pulse is presented and comparison with coincidence
scheme is given.
\end{abstract}

\section{Introduction}

There is a project to measure directly n-n collision for checking
charge symmetry of nuclear forces~\cite{1}. It is accepted that
the best neutron source to perform such measurements is the
Russian pulsed YAGUAR reactor. Some preliminary measurements and
numerical simulations for expected experimental geometry had been
performed~\cite{2}. We want to show here an analytical approach to
calculations. First we obtain analytical momentum spectrum of
scattered neutrons, then the time of flight spectrum of neutrons
detected by a single counter. After that we consider coincidence
scheme where we have two detectors, and calculate time of flight
spectrum for one detector and delay time spectrum for the second
one. We considered coincidence scheme because from the very
beginning of discussions about the project, and all the time
during preparation of the experiment, many people continue to
express the opinion that the coincidence scheme has an advantage
comparing to the single detector measurement. They claim that loss
of intensity, which they usually estimated at the level of 20\%,
will be surpassed by much higher suppression of background. We
show here analytically that in the coincidence scheme effect is so
much suppressed, that the question about the background level
becomes irrelevant.
\section{Estimation of the effect}

The scheme of the experiment is presented in Fig. 1 borrowed
from~\cite{1}. The YAGUAR reactor 1 gives a pulse of length
$t_p=0.68$ ms, during which a huge amount of neutrons with flux
density $\Phi=0.77\times10^{18}$ n/cm$^2$s is released. After a
moderator at room temperature $T$ neutrons in the thermal
Maxwellian spectrum arrive at the volume 2 ($V=1.13$ cm$^3$),
where they collide with each other and some of them after
collision fly along the neutron guide 3 with collimators 4, and
arrive at the detector 5, where they are registered with
$\sim100$\% efficiency. The collimators 4 determine the solid
angle $\Delta\Omega=0.64\times10^{-4}$, at which the volume $V$ is
visible by the detector. The estimated number of neutrons that can
be registered at a single pulse is equal to
\begin{equation}\label{e1}
N_e=2n^2Vt_pv_T|b|^2d\Omega,
\end{equation}
where factor 2 takes into account that the detector can register
scattered neutron or neutron-scatterer. The square of the
scattering amplitude $|b|^2$ is defined as: $|b|^2=|b_0|^2/4$,
where $b_0$ is the singlet scattering amplitude, which is accepted
to be $18$ fm, and factor $1/4$ is statistical weight of the
singlet scattering. Therefore $|b|^2=8.1\times10^{-25}$ cm$^2$.
The speed $v_T$ corresponds to the thermal speed $v_T=2200$ m/s,
and the factor $v_T|b|^2$ determines number of collisions in the
neutron gas per unit time. The factor $n^2$ is the square of the
neutron density: $n=\Phi/v_T=3\times10^{12}$ cm$^{-3}$. After
substitution of all the parameters into (\ref{e1}) we find
$N_e\approx170$ neutrons per pulse. However it is the estimation
number. To find real number counted by the single detector, $N_s$,
it is necessary to calculate the scattering process. Calculation
shows that $N_s=FN_e$, where factor $F$ is of the order unity.
Monte Carlo calculations in~\cite{1} give $F=0.83$. Analytical
calculations presented below give $F=0.705$.
    The number of neutrons per pulse counted at coincidence, if the
neutrons trap 6 is replaced by another detector, can be estimated
as
\begin{equation}\label{e2}
N_{ec}=N_sd\Omega \tau/t_T,
\end{equation}
where $\tau$ is the width of the coincidence window, $t_T=L/v_T$
is the average length of measurement time after the reactor pulse,
and $L\approx 12$ m is the average distance between collision
volume and the detectors. In the experimental scheme of Fig. 1 the
time $t_T$ is of the order 5 ms. If we accept $\tau\approx
t_p=0.5$ ms, then the ratio $\tau/t_T$ is 0.1. The factor
$d\Omega$ is included in (\ref{e2}), because only neutrons in this
solid angle will be registered by the second detector. The total
factor, which suppresses the estimated number of neutrons
registered per single pulse in coincidence scheme, is of the order
$10^{-5}$, therefore the estimated number of counts in coincidence
scheme will be $10^{-3}$, so the experiment becomes non feasible,
and the level of the background, which is determined by neutron
scattering on the residual gas atoms present at even very good vacuum
conditions, becomes irrelevant. The analytical calculations,
presented below, show that the real number of counted neutrons in
coincidence scheme contains even additional small factor
$F_c=0.15$.

\begin{figure}[t]
{\par\centering\resizebox*{6cm}{!}{\includegraphics{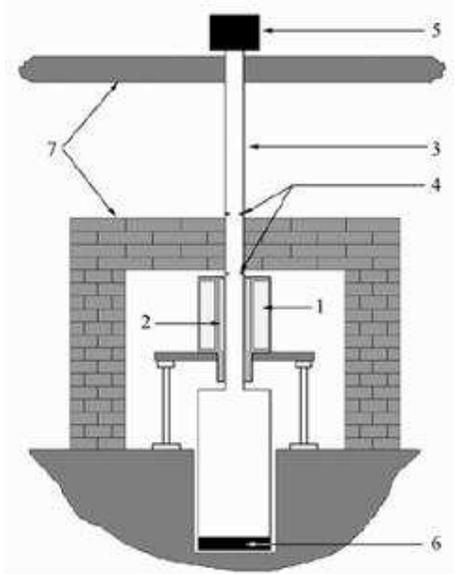}}\par}
\caption{Scheme of the experiment on direct measurement of n-n
scattering~\cite{1}. 1 --- reactor core; 2 --- volume of
collisions; 3 --- neutron guide; 4 --- collimators; 5 ---
detector; 6 --- neutrons trap.} \label{f1}
\end{figure}

\section{The analytical calculation of neutron scattering in the thermal neutron gas}

Our calculations will be based on the standard scattering theory
of neutron scattering in the atomic gas. Our main feature is that
we shall make calculations directly in the laboratory reference
frame without transition to the center of mass system. First we
remind all the definitions of the standard scattering theory and
then present analytical calculations of all the required
integrals.

\subsection{The standard scattering theory}

The standard scattering theory starts with the Fermi golden rule,
according to which one can write down the probability of the
neutron scattering per unit time on an arbitrary system as
\begin{equation}\label{wpg1a}
dw(\kk_i\to\kk_f,\lambda_i\to\lambda_f)=\frac{2\pi}{\hbar}\left|\langle\lambda_f,\kk_f|U|\lambda_i,
\kk_i\rangle\right|^2\delta(E_{fk}+E_{f\lambda}-E_{ik}-E_{i\lambda})\rho(E_{fk}),
\end{equation}
where $|\kk_i>$ , $|\lambda_i>$ are initial, $|\kk_f>$,
$|\lambda_f>$ are final states of the neutron and system with
energies $E_{ik}$, $E_{i\lambda}$, $E_{fk}$, $E_{f\lambda}$
respectively, $U$ is the neutron-system interaction potential,
which in the neutron atom scattering is accepted in the form of
the Fermi pseudo potential
\begin{equation}
U=\frac{\hbar^2}{2m}4\pi b\delta(\rr_1-\rr_2). \label{wpg1d}
\end{equation}
Here $\rr_1$, $\rr_2$ are positions of the neutron and the system,
$\rho(E_{fk})$ is the density of the neutron final states
\begin{equation}
\rho(E_k)=\left(\frac{L}{2\pi}\right)^3d^3k, \label{wpg1b}
\end{equation}
$E_k=\hbar^2k^2/2m$, $m$ is the neutron mass, and $L$ is the size
of some arbitrary space cell.

We suppose that the system is an atom with  mass $M=m$, and
momentum $\p$. The initial and final states of the neutron and
atom are described with similar wave functions
\begin{equation}
|\kk_{i,f}>=\frac{1}{L^{3/2}}\exp(i\kk_{i,f}\rr),\qquad
|\lambda_{i,f}>\equiv|\p_{i,f}>=\frac{1}{L^{3/2}}\exp(i\p_{i,f}\rr),
\label{m1}
\end{equation}
where $\kk_{i,f}$ and $\p_{i,f}$ are initial and final neutron and
atom momenta respectively.

The flux density of the single incident neutron is
\begin{equation}\label{me0}
j_i=\hbar k_i/mL^3.
\end{equation}
The scattering cross section at the given initial and final states
is the ratio
\begin{equation}\label{me0a}
d\sigma(\kk_i\to\kk_f,\p_i\to\p_f)=\frac1{j_i}dw(\kk_i\to\kk_f,\p_i\to\p_f).
\end{equation}
At the next step we need to sum this cross section over final
states of the system and average over initial states. In our case
summation over the system final states is the integration over
density of the atomic final states
\begin{equation}
\rho(E_{pf})=\left(\frac{L}{2\pi}\right)^3d^3p_f. \label{wpg1b1}
\end{equation}
This integration gives the cross section for the given initial
states as
$$d\sigma(\kk_i\to\kk_f,\p_i)$$
\begin{equation}\label{me0a2a}
=2\frac{2\pi m}{\hbar^2 k_i}\frac{L^9d^3k_f}{(2\pi)^6}\int d^3p_f
\left|\langle\p_f,\kk_f|U|\p_i,
\kk_i\rangle\right|^2\delta(E_{fk}+E_{fp}-E_{ik}-E_{ip}),
\end{equation}
where $E_p=\hbar^2p^2/2M$, $E_k=\hbar^2k^2/2m$, and the additional
factor 2 means that the atom and neutron are the same particles,
therefore we can detect with the same probability the scattered
neutron in the phase element $d^3k_f$ or an atom in the element
$d^3p_f$.

For our experiment we need not a cross section, but the number of
the neutrons $dN(\kk_i,\p_i,\kk_f)$ scattered in the element
$d^3k_f$. This number is determined by the number of collisions of
neutrons with atoms, so the number of scattered neutrons is equal
to
\begin{equation}\label{q}
dN(\kk_i,\p_i,\kk_f)=dn_a(p_i)dn_n(k_i)
Vdt_pvd\sigma(p_i,\kk_i\to\kk_f),
\end{equation}
where $dn_a(p_i)$, $dn_n(k_i)$ are the number densities of atoms
and neutrons with initial momenta $\p_i$ and $\kk_i$ respectively,
$v=\hbar|\p_i-\kk_i|/m$ is the relative neutron-atom velocity, and
$V$, $dt_p$ are volume and time, where
collisions create detectable neutrons.

Since our atoms and neutrons have the same Maxwellian distribution
with the temperature $T$, the densities $dn_a(p_i)$ and
$dn_n(k_i)$ are
\begin{equation}
dn_a(\q)=dn_n(\q)=n\frac{d^3q}{(2\pi
T)^{3/2}}\exp\left(-\frac{q^2}{2T}\right),\label{wpg1b2}
\end{equation}
where $n$ is the average neutrons density, the letter $T$ denotes
reduced temperature $T=mk_B[T]/\hbar^2$, and $[T]$ is the
temperature in Kelvin degrees. To find the total number of
neutrons $dN(\kk_f)$ scattered into element $d^3k_f$ of the final
momentum space we must integrate (\ref{q}) over
$dn_a(p_i)dn_n(k_i)$, after which we get
$$dN(\kk_f)=2n^2Vdt_p\frac{1}{(2\pi T)^3}\frac{L^9d^3k_f}{(2\pi)^6}\frac{2\pi}{\hbar}
\frac{2m}{\hbar^2} \int d^3k_i\int
d^3p_i\frac{|\p_i-\kk_i|}{k_i}$$
\begin{equation}\label{me0a2a1}
\times\exp\left(-\frac{p_i^2+k_i^2}{2T}\right)\int d^3p_f
\left|\langle\p_f,\kk_f|V|\p_i,
\kk_i\rangle\right|^2\delta(k_f^2+p_f^2-k_i^2-p_i^2).
\end{equation}
The matrix element of the potential (\ref{wpg1d}) is
\begin{equation}\label{me}
\langle\p_f,\kk_f|V|\p_i, \kk_i\rangle=4\pi
b\frac{\hbar^2}{2m}\frac{(2\pi)^3}{L^6}\delta(\p_i+\kk_i-\p_f-\kk_f),
\end{equation}
and its square is
\begin{equation}\label{me1}
\left|\langle\p_f,\kk_f|V|\p_i, \kk_i\rangle\right|^2=|4\pi
b|^2\left(\frac{\hbar^2}{2m}\right)^2\frac{(2\pi)^3}{L^9}\delta(\p_i+\kk_i-\p_f-\kk_f).
\end{equation}
After substitution of (\ref{me1}) into (\ref{me0a2a1}) we can
extract $|b|^2$ from the square of the matrix
element, $d\Omega$ from $d^3k_f$ and introduce the thermal speed
$v_T=\hbar\sqrt{2T}/m$. As a result we obtain
\begin{equation}\label{e3}
dN(\kk_f)=N_eg(\kk_f)\frac{dk_f}{\sqrt{2T}},
\end{equation}
where $N_e$ is given in (\ref{e1}), and $g(\kk_f)$ is
$$g(\kk_f)=\frac{2}{\pi^3}\frac{k^2_f}{(2T)^{3}}
\int d^3k_i\int d^3p_i\frac{|\p_i-\kk_i|}{k_i}$$
\begin{equation}\label{me2}
\times\int d^3p_f\exp\left(-\frac{p_f^2+k_f^2}{2T}\right)
\delta(\p_i+\kk_i-\p_f-\kk_f)\delta(k_f^2+p_f^2-k_i^2- p_i^2).
\end{equation}

Integration over $d^3p_i$ gives
\begin{equation}\label{me2a}
g(\kk_f)=\frac{2}{\pi^3}\frac{k^2_f}{(2T)^{3}} \int
d^3p_f\exp\left(-\frac{p_f^2+k_f^2}{2T}\right) \int
d^3k_i\frac{|\PP-2\kk_i|}{k_i}\delta(k_f^2+p_f^2-k_i^2-(\PP-\kk_i)^2),
\end{equation}
where $\PP=\p_f+\kk_f$ is the total momentum of two particles.

With all these definitions in hands we can directly calculate the
spectrum of scattered neutrons

\subsection{Analytical calculation of the integrals}

First we calculate the integral
$$Q(\kk_f,\p_f)=\int d^3k_i
\frac{|\PP-2\kk_i|}{k_i}\delta(k_f^2+p_f^2-k_i^2- (\PP-\kk_i)^2)
$$
\begin{equation}\label{me7}
=2\int \frac{d^3k_i}{k_i} |2\kk_i-\PP|\delta((\kk_f-\p_f)^2+
(2\kk_i-\PP)^2).
\end{equation}
After change of variables $2\kk_i-\PP=\uu$ we obtain
\begin{equation}\label{me8}
Q(\kk_f,\p_f)=\frac12\int u \frac{d^3u}{|\uu+\PP|}\delta(u^2-q^2),
\end{equation}
where $q^2=(\kk_f-\p_f)^2$.

After representation $ud^3u=(u^2du^2/2)d\varphi d\cos\theta$,
where polar axis is chosen along the vector $\PP$, we can
integrate over $d\varphi$ and $d(u^2)$. As a result we get
\begin{equation}\label{me9}
Q(\kk_f,\p_f)=\int\limits_{-1}^1\frac{\pi
q^2d\cos\theta}{2\sqrt{q^2+2Pq\cos\theta+P^2}}.
\end{equation}
Integration over $d\cos\theta$ gives
\begin{equation}\label{me99}
Q(\kk_f,\p_f)=\frac{\pi q}{2P}(q+P-|q-P|).
\end{equation}
The last factor is equal to $2q$, if $q<P$, and it is equal to
$2P$, if $q>P$. Which one of these inequalities is satisfied
depends on the angle $\theta_f$ between vectors $\kk_f$ and
$\p_f$. Inequality $q<P$ is satisfied, when $\cos\theta_f>0$, and
inequality $q>P$ is satisfied, when $\cos\theta_f<0$. Therefore
Eq. (\ref{me99}) is representable in the form
\begin{equation}\label{me10}
Q(\kk_f,\p_f)= \pi
q\left(\Theta(\cos\theta_f<0)+\Theta(\cos\theta_f>0)
\frac{q}{P}\right),
\end{equation}
where $\Theta(x)$ is the step function equal to unity, when
inequality in its argument is satisfied, and to zero in the
opposite case.

\subsection{The spectrum of neutrons, counted by a single detector}

Substitution of (\ref{me10}) into (\ref{me2a}) gives
\begin{equation}\label{e4}
g(\kk_f)=\int d^3p_fw(\kk_f,\p_f),
\end{equation}
where
\begin{equation}\label{e5}
w(\kk_f,\p_f)=\frac{2}{\pi^3}\frac{k^2_f}{(2T)^{3}}
\exp\left(-\frac{p_f^2+k_f^2}{2T}\right)Q(\kk_f,\p_f).
\end{equation}
To obtain spectrum of neutrons counted by a single detector we
represent $d^3p_f=p^2_fdp_fd\Omega_f$, and integrate $Q(\kk,\p)$
over $d\Omega_f$.  As a result we obtain (in the following we omit
subscripts $f$ of variables)
$$I(k,p)=\int Q(\kk,\p)d\Omega=
2\pi^2\left(\int_{-1}^0
d\cos\theta|\kk-\p|+\int_0^1d\cos\theta\frac{(\kk-\p)^2}{|\kk+\p|}\right)$$
$$=2\pi^2\left(\int_{0}^1
d\cos\theta|\kk+\p|+\int_0^1d\cos\theta\left[\frac{2(k^2+
p^2)}{|\kk+\p|}-|\kk+\p|\right]\right)$$
\begin{equation}\label{me11a}
=\frac{(2\pi)^2}{pk}(p^2+k^2)(p+k-\sqrt{p^2+k^2}).
\end{equation}

Substitution of (\ref{me11a}) into (\ref{e5}) and change of
variables $x=p/k$, $y=k/\sqrt{2T}$ gives
\begin{equation}\label{nn2}
g(\kk_f)\equiv f(y)=\frac2\pi \exp(-y^2)y^2J(y),
\end{equation}
where
\begin{equation}\label{me15}
J(y)=2y^4\int_0^\infty
2xdx\exp(-x^2y^2)(1+x^2)\left[(x+1)-\sqrt{x^2+1}\right].
\end{equation}
Integration by parts gives
\begin{equation}\label{me15a}
J(y)=2y^2\int_0^\infty
dx\exp(-x^2y^2)(1+2x+3x^2-3x\sqrt{x^2+1})=y\sqrt\pi+J_1(y),
\end{equation}
where
$$J_1(y)=2y^2\int_0^\infty
xdx\exp(-x^2y^2)(2+3x-3\sqrt{x^2+1})$$
\begin{equation}\label{me15b}
=-1+3\int_0^\infty
xdx\exp(-x^2y^2)(1-\frac{x}{\sqrt{x^2+1}})=-1+3\frac{\sqrt\pi}{2y}\{1-e^{y^2}[1-\Phi(y)]\},
\end{equation}
and
\begin{equation}\label{nn12}
\Phi(y)=\frac2{\sqrt\pi}\int_0^ydx\exp(-x^2).
\end{equation}
Substitution of (\ref{me15b}) into (\ref{me15a}) gives
\begin{equation}\label{me15c}
J(y)=y\sqrt\pi-1+3\frac{\sqrt\pi}{2y}\{1-e^{y^2}[1-\Phi(y)]\}.
\end{equation}
The momentum spectrum $f(y)$ from Eq. (\ref{nn2}) with account of
(\ref{me15c}) is shown in Fig. 2. Numerical integration of this
function gives $F=\int_0^\infty f(y)dy=0.705$.

\begin{figure}[t]
{\par\centering\resizebox*{6cm}{!}{\includegraphics{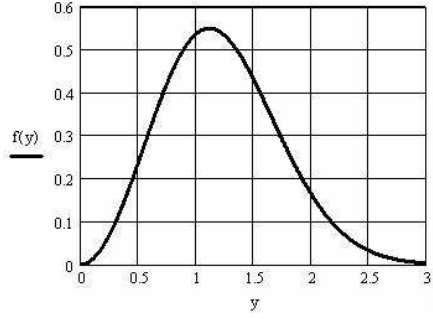}}\par}
\caption{Spectrum $y^2F(y)\exp(-y^2)$ of the neutrons detected by
a single detector in dimensionless units $y=v/v_T$.} \label{f2}
\end{figure}

\subsection{Time of flight spectrum of a single detector}

In the experiment the time of flight (TOF) spectrum is measured.
To transform (\ref{nn2}) into TOF spectrum we multiply it by unity
\begin{equation}\label{u}
1=dt\delta(t-L/v_Ty),
\end{equation}
where $L$ is the distance between scattering volume and the
detector, and integrate over $dy$. After that we obtain
\begin{equation}\label{me19e}
\dot N_s(y(t))=f(L/v_Tt)\frac{L}{v_Tt^2}=f(y)\frac{y}{t}.
\end{equation}

\section{Registration by two detectors in coincidence}

Let's consider the case, when neutrons are registered in
coincidence by two detectors on the opposite sides of the
collision volume. It means that the angle between $\kk_f$ and
$\p_f$ is approximately 180$^\circ$. Since we register both
neutrons, we should not integrate (\ref{me10}) over $d^3p_f$.
Instead we should accept $\kk_f\p_f<0$, $\p_f\approx-\kk_f$, and
$d^3p_f=p^2_fdp_fd\Omega$ with the same $d\Omega$ as in $d^3k_f$.
Taking into account Eq. (\ref{e3}), (\ref{e4}) and (\ref{e5}) we
can represent the number of neutrons counted by two detectors as
\begin{equation}\label{me2z}
dN(\kk_f,\p_f)=N_ed\Omega_{pf}\frac{2}{\pi^2}\frac{k^2_fdk_f}{(2T)^{7/2}}
qp^2_fdp_f\exp\left(-\frac{p_f^2+k_f^2}{2T}\right).
\end{equation}
After transformation to dimensionless variables $y=k_f/\sqrt{2T}$
and $z=p_f/\sqrt{2T}$ we get
\begin{equation}\label{me2yy}
dN(\kk_f,\p_f)=N_ed\Omega_{p}G(y,z)dydz,
\end{equation}
where
\begin{equation}\label{me2x}
G(y,z)=\frac{2y^2z^2}{\pi^2}(y+z) \exp(-y^2-z^2),
\end{equation}
and we replaced $q$ by $k_f+p_f$.

To get TOF spectrum in one detector and coincidence count in the
second one with coincidence window $\tau$ we must multiply
(\ref{me2yy}) by the unit
\begin{equation}\label{un}
1=dt\delta(t-L/v_Ty)dt'\delta(t'-L/v_Tz+t)
\end{equation}
and integrate over $dydz$. As a result we obtain
\begin{equation}\label{me2y}
\dot N_c\equiv
dN(\kk_f,\p_f)/dt=N_0d\Omega_{p}G\left(\frac{L}{v_Tt},\frac{L}{v_T(t+t')}\right)\frac{dt'
L^2}{v_T^2t^2(t+t')^2}.
\end{equation}
After integration over $dt'$ in the range of the coincidence
window $\tau$ we can put $z\approx y$, and finally get
\begin{equation}\label{me2u}
\dot N_c\approx
N_0d\Omega_{p}\frac{4y^7}{\pi^2}\exp(-2y^2)\frac{\tau}{t^2}.
\end{equation}
For comparison of TOF spectrum of two and single detectors it is
useful to find ratio of (\ref{me2u}) to (\ref{me19e}). This ratio
is
\begin{equation}\label{R}
W=\frac{\dot N_c}{\dot N_s}= d\Omega_{p}\frac{\tau}{t}R(y),
\end{equation}
\begin{figure}[t]
{\par\centering\resizebox*{6cm}{!}{\includegraphics{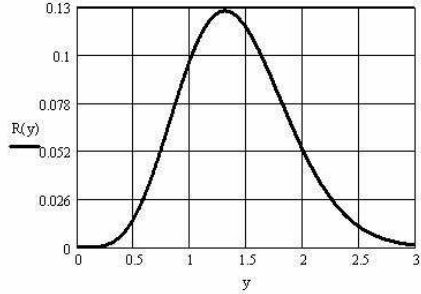}}\par}
\caption{Dependence of $R(y)$ on $y=k/\sqrt{2mT}$.} \label{f3}
\end{figure}
where
\begin{equation}\label{R1}
R(y)=\frac{4y^6}{\pi^2f(y)}\exp(-2y^2).
\end{equation}
The function $R(y)$ is shown in Fig. 3. Its integral $\int dyR(y)$
is equal to 0.15. So we can tell that the ratio is approximately
\begin{equation}\label{R11}
W\approx 0.1d\Omega\frac{\tau}{t},
\end{equation}
as is said in section 2.

\section{Conclusion}

We have shown that the effect of n-n scattering experiment and
spectrum of detected neutrons in a single detector can be
calculated analytically with the standard scattering theory
without transformation to center of mass system. Analytically
calculated factor $F=0.705$ is close to that $F=0.83$, calculated
by Monte Carlo method. The difference can be attributed to
slightly different spectra of neutrons in the collision volume. In
Monte Carlo calculations spectrum contained Maxwellian part and
epithermal tail, while for analytical calculations we used only
Maxwellian part. We did not calculated background which is related
to scattering of neutrons on gas molecules, but we claim that it
also can be calculated analytically. One of the main conclusions
of this paper is that coincidence scheme for this type of
experiment is absolutely impractical, because the effect becomes
so low, that the level of the background becomes irrelevant.

\section*{acknowledgments} I am grateful to A.V.Strelkov for
introduction me to this problem and to W.Furman, E.Lychagin and
A.Muzychka for interest and discussion.

\section{The history of submissions and rejections}

I submitted this paper on March 10 to the same journal J. Phys. G:
Nucl. Part. Phys., where the two papers\cite{1,2} were published.
On March 31 I received the electronic mail from editors with
subject {\bf  Final decision on your article from J. Phys. G:
Nucl. Part. Phys.}, which meant that no more negotiations are
supposed. The letter contained the referee report. I am not permitted by arXiv
policy to present full content of the report, so I give a paraphrase of it.

\subsection{In his report the referee writes}
that the paper is not worth of further consideration by the journal,
because the journal has already published articles where neutron spectra were calculated by
Monte Carlo techniques. Analytical calculations by standard technique are not worth to
be published. A new result of the paper, which shows
that coincidence scheme is not profitable was also obtained by participants of the project
(their result was not yet
published or submitted for publication), therefore it is necessary to reject the paper.

\subsection{My remark not sent to the editor}

In my opinion analytical calculation is much superior to Monte
Carlo one. If it was not known and became known only later, it is
worth its own publication. Even if it gives the same result as
Monte Carlo one, it is very important, because it proves that
Monte Carlo calculations was made correctly. No one can check
Monte Carlo calculations. Every one can check analytical
calculations. It is great that for analytical calculations one
need nothing except standard theory, because, if it were necessary
to make some new assumption, the merit of analytical calculation
would become doubtful.

Referee also pointed out, that the papers [1,2] contain calculation of signal and background,
while I calculated only signal. To this remark I can say that I am able to calculate the background too.
However this analytical calculation is more laborious and why to do it, if it had already
been calculated by the Monte-Carlo method, and analytical calculation are not
worth of a ``stand-alone'' publication?

After getting this reply I reconsidered my article and improved
it. But in essence it remained the same. So the referee, if he were
able to read the improved version, would have no reason to change
his report.

\subsection{The end of the story}

On April 3 I submitted the paper to Yad.Phys. (Russian Nuclear
Physics), and on April 29 I received the referee report, which
approved the paper and contained some comments that helped me to
improve it even further. So I want to express my gratitude to him.

\subsection{The story continues}

After correction of the article I submitted it again to Yad. Phys.
but after few days I obtained the letter, where editors asked me
to delete section 6. I replied why? Whether it is not my right to
publish everything, which is related to the problem? More over the
referee did not require to omit this part. But in the next letter
the editors informed me that the referee considered the deletion
of the section 6 as self-evident. I sent to the editorial board my
arguments why to accept my paper with the section 6. They are:
\begin{enumerate}
\item It is nasty to forbid something which can be permitted.
\item It is nasty to apply power where it is useless. \item
Publication of the section 6 is harmful only for editorial board
of the J. Phys. G, and their referee, but is very profitable for
the whole scientific community, because it shows that
irresponsible referee reports will be published and it is a real
punishment for them. \item Any decision of the Yad.
Phys. editors will be historical one, because I shall publish everything
in the ArXiv, but the positive decision will demonstrate that the
editorial board agree with me and takes the responsibility for
future reports of its referees. \item The section 6 is not
irrelevant to the content of the paper. It rises the important
question: whether analytical calculations merit publication as a
stand-alone article or not, if everything can be calculated
numerically by, say, Monte-Carlo method.
\end{enumerate}
Notwithstanding of my arguments the verdict was -- to delete the
section 6. One of the vice chief editors (he is not anonymous and
advised me not to reveal his name) wrote me that he CANNOT publish
because it contradicts to the LAW of GENRE. I asked him which
article of the GENRE LAW does he refers? One of the Journals
publishes my articles with referee
reports. The editorial board does not think that it contradicts
the GENRE LAW and I take off my hat to their editors. However,
since there are no arguments except the claim that the Journal
CANNOT publish the section 6, I consider it as a demonstration of
power to which I am to obey. I delete the section, but I shall
publish the full article in the ArXiv, and there I shall explain
why the section 6 is excluded from the article published in
Yad.Phys.

\subsection{An attempt to publish JINR preprint}

I wanted also to publish my paper with section 6 as a
preprint of JINR. The chief of publishing department told me that
she has no right to publish such a section without approval by Scientific Secretary (SS).
I applied to SS for permission, but in
vain. Then I appealed to our director (D) with the same arguments
as above.

-Stop! --- said one of my friends to whom I told the story. --- I can predict the end of it.
The D will sent your article again to SS. The SS will return it to D with a note: ``I consider it
not appropriate'', and the D will write ``I agree'' and sign such a resolution!

I was astonished how smart was my friend!

\end{document}